\documentclass[journal]{IEEEtran}
\usepackage{blindtext}
\usepackage{graphicx}
\usepackage{booktabs} 
\usepackage{multirow} 
\usepackage{color}

\ifCLASSINFOpdf
\else
\fi

\usepackage[cmex10]{amsmath}
\usepackage{array}
\usepackage[tight,footnotesize]{subfigure}
\usepackage[font=footnotesize]{subfig}

\begin{document}
%
\title{Single-Channel Multi-talker Speech Recognition with Permutation Invariant Training}
%
%
%

\author{Yanmin Qian,~\IEEEmembership{Member,~IEEE,}
        Xuankai Chang,~\IEEEmembership{Student Member,~IEEE,}
        and Dong Yu,~\IEEEmembership{Senior Member,~IEEE}
\thanks{Yanmin Qian and Xuankai Chang are with Computer Science and Engineering Department, Shanghai Jiao Tong University, Shanghai, 200240 P. R. China (\{yanminqian,xuank\}@sjtu.edu.cn).}
\thanks{Dong Yu is with Tencent AI Lab, Seattle, USA (dyu@tencent.com).}
}

\markboth{IEEE/ACM TRANSACTIONS ON AUDIO, SPEECH, AND LANGUAGE PROCESSING}
{Qian \MakeLowercase{\textit{et al.}}: Single-Channel Multi-talker Speech Recognition with Permutation Invariant Training}

\maketitle

\begin{abstract}
Although great progresses have been made in automatic speech recognition (ASR), significant performance degradation is still observed when recognizing multi-talker mixed speech. In this paper, we propose and evaluate several architectures to address this problem under the assumption that only a single channel of mixed signal is available. Our technique extends permutation invariant training (PIT) by introducing the front-end feature separation module with the minimum mean square error (MSE) criterion and the back-end recognition module with the minimum cross entropy (CE) criterion. More specifically, during training we compute the average MSE or CE over the whole utterance for each possible utterance-level output-target assignment, pick the one with the minimum MSE or CE, and optimize for that assignment. This strategy elegantly solves the label permutation problem observed in the deep learning based multi-talker mixed speech separation and recognition systems. The proposed architectures are evaluated and compared on an artificially mixed AMI dataset with both two- and three-talker mixed speech. The experimental results indicate that our proposed architectures can cut the word error rate (WER) by 45.0\% and 25.0\% relatively against the state-of-the-art single-talker speech recognition system across all speakers when their energies are comparable, for two- and three-talker mixed speech, respectively. To our knowledge, this is the first work on the multi-talker mixed speech recognition on the challenging speaker-independent spontaneous large vocabulary continuous speech task.

\end{abstract}

\begin{IEEEkeywords}
permutation invariant training, multi-talker mixed speech recognition, feature separation, joint-optimization
\end{IEEEkeywords}

%
\IEEEpeerreviewmaketitle

\section{Introduction} \label{sec:intro}

Thanks to the significant progresses made in the recent years \cite{ASRBook-Yu2014,PretrainVSFineTune-Yu2010,CD-DNN-HMM-dahl2012,CD-DNN-HMM-SWB-seide2011,DNN4ASR-hinton2012,CNN4ASR-Abdel-Hamid2012,CNN-Trans-Abdel-Hamid2014,CLDNN-sainath2015,DeepCNN-bi2015,DeepCNN-qian2016.1,DeepCNN-qian2016.2,TFCNN-mitra2015,TDNN-peddinti2015,VGG-secru2016,Deepspeech2-amodei2015,FSMN-zhang2015,LACE-yu2016,HumanParity-Xiong2016,PIT-yu2017,PIT-Kolbak2017}, the ASR systems now surpassed the threshold for adoption in many real-world scenarios and enabled services such as Microsoft Cortana, Apple's Siri and Google Now, where close-talk microphones are commonly used.

However, the current ASR systems still perform poorly when far-field microphones are used. This is because many difficulties hidden by close-talk microphones now surface under distant recognition scenarios. For example, the signal to noise ratio (SNR) between the target speaker and the interfering noises is much lower than that when close-talk microphones are used. As a result, the interfering signals, such as background noise, reverberation, and speech from other talkers, become so distinct that they can no longer be ignored.

In this paper, we aims at solving the speech recognition problem when multiple talkers speak at the same time and only a single channel of mixed speech is available. Many attempts have been made to attack this problem. Before the deep learning era, the most famous and effective model is the factorial GMM-HMM \cite{FactorialHMM-ghahramani1997}, which outperformed human in the 2006 monaural speech separation and recognition challenge \cite{MonauralSpeechSepChallenge-Cooke2010}. The factorial GMM-HMM, however, requires the test speakers to be seen during training so that the interactions between them can be properly modeled. Recently, several deep learning based techniques have been proposed to solve this problem \cite{PIT-yu2017,PIT-Kolbak2017,SingleChannelSep-Weng2015,DeepClustering-hershey2015,DeepClustering2-isik2016,AtrractorNet4SpeechSeparation-chen2017}. The core issue that these techniques try to address is the label ambiguity or permutation problem (refer to Section \ref{sec:pit} for details).

In Weng et al. \cite{SingleChannelSep-Weng2015} a deep learning model was developed to recognize the mixed speech directly. To solve the label ambiguity problem, Weng et al. assigned the senone labels of the talker with higher instantaneous energy to output one and the other to output two. This, although addresses the label ambiguity problem, causes frequent speaker switch across frames. To deal with the speaker switch problem, a two-speaker joint-decoder with a speaker switching penalty was used to trace speakers. This approach has two limitations. First, energy, which is manually picked, may not be the best information to assign labels under all conditions. Second, the frame switching problem introduces burden to the decoder. 

In Hershey et al. \cite{DeepClustering-hershey2015,DeepClustering2-isik2016} the multi-talker mixed speech is first separated into multiple streams. An ASR engine is then applied to these streams independently to recognize speech. To separate the speech streams, they proposed a technique called deep clustering (DPCL). They assume that each time-frequency bin belongs to only one speaker and can be mapped into a shared embedding space. The model is optimized so that in the embedding space the time-frequency bins belong to the same speaker are closer and those of different speakers are farther away. During evaluation, a clustering algorithm is used upon embeddings to generate a partition of the time-frequency bins first, separated audio streams are then reconstructed based on the partition. In this approach, the speech separation and recognition are usually two separate components. 

Chen et al. \cite{AtrractorNet4SpeechSeparation-chen2017} proposed a similar technique called deep attractor network (DANet). Following DPCL, their approach also learns a high-dimensional embedding of the acoustic signals. Different from DPCL, however, it creates cluster centers, called attractor points, in the embedding space to pull together the time-frequency bins corresponding to the same source. The main limitation of DANet is the requirement to estimate attractor points during evaluation time and to form frequency-bin clusters based on these points.

In Yu et al. \cite{PIT-yu2017} and Kolbak et al.\cite{PIT-Kolbak2017}, a simpler yet equally effective technique named permutation invariant training (PIT)\footnote{In \cite{DeepClustering-hershey2015}, a similar permutation free technique, which is equivalent to PIT when there are exactly two-speakers, was evaluated with negative results and conclusion.} was proposed to attack the speaker independent multi-talker speech separation problem. In PIT, the source targets are treated as a set (i.e., order is irrelevant). During training, PIT first determines the output-target assignment with the minimum error at the utterance level based on the forward-pass result. It then minimizes the error given the assignment. This strategy elegantly solved the label permutation problem. However, in these original works PIT was used to separate speech streams from mixed speech. For this reason, a frequency-bin mask was first estimated and then used to reconstruct each stream. The minimum mean square error (MMSE) between the true and reconstructed speech streams was used as the criterion to optimize model parameters.

Moreover, most of previous works on multi-talker speech still focus on speech separation \cite{PIT-yu2017,PIT-Kolbak2017,DeepClustering-hershey2015,DeepClustering2-isik2016,AtrractorNet4SpeechSeparation-chen2017}. In contrast, the multi-talker speech recognition is much harder and the related work is less. There has been some attempts, but the related tasks are relatively simple. For example, the 2006 monaural speech separation and recognition challenge \cite{FactorialHMM-ghahramani1997,MonauralSpeechSepChallenge-Cooke2010,SingleChannelSep-Weng2015,HERSHEY201045,rennie2010single} was defined on a speaker-dependent, small vocabulary, constrained language model setup, while in \cite{DeepClustering2-isik2016} a small vocabulary reading style corpus was used. We are not aware of any extensive research work on the more real, speaker-independent, spontaneous large vocabulary continuous speech recognition (LVCSR) on multi-talker mixed speech before our work.


In this paper, we attack the multi-talker mixed speech recognition problem with a focus on the speaker-independent setup given just a single-channel of the mixed speech. Different from \cite{PIT-yu2017,PIT-Kolbak2017}, here we extend and redefine PIT over log filter bank features and/or senone posteriors. In some architectures PIT is defined upon the minimum mean square error (MSE) between the true and estimated individual speaker features to separate speech at the feature level (called PIT-MSE from now on). In some other architectures, PIT is defined upon the cross entropy (CE) between the true and estimated senone posterior probabilities to recognize multiple streams of speech directly (called PIT-CE from now on). Moreover, the PIT-MSE based front-end feature separation can be combined with the PIT-CE based back-end recognition in a joint optimization architecture. We evaluate our architectures on the artificially generated AMI data with both two- and three-talker mixed speech. The experimental results demonstrate that our proposed architectures are very promising.

The rest of the paper is organized as follows. In Section \ref{sec:problem} we describe the speaker independent multi-talker mixed speech recognition problem. In Section \ref{sec:pit} we propose several PIT-based architectures to recognize multi-streams of speech. We report experimental results in Section \ref{sec:exp} and conclude the paper in Section \ref{sec:conclusion}.

\section{Single-Channel Multi-Talker Speech Recognition} \label{sec:problem}

In this paper, we assume that a linearly mixed single-microphone signal ${y}[n] = \sum_{s=1}^{S} {x}_s[n]$ is known, where ${x}_s[n], s=1, \cdots, S$ are $S$ streams of speech sources from different speakers. Our goal is to separate these streams and recognize every single one of them. In other words, the model needs to generate $S$ output streams, one for each source, at every time step. However, given only the mixed speech ${y}[n]$, the problem of recognizing all streams is under-determined because there are an infinite number of possible ${x}_s[n]$ (and thus recognition results) combinations that lead to the same ${y}[n]$. Fortunately, speech is not random signal. It has patterns that we may learn from a training set of pairs $\mathbf{y}$ and ${\mathbf{\ell}^s,s=1,\cdots,S}$, where $\mathbf{\ell}^s$ is the senone label sequence for stream $s$. 

In the single speaker case, i.e., $S=1$, the learning problem is significantly simplified because there is only one possible recognition result, thus it can be casted as a simple supervised optimization problem. Given the input to the model, which is some feature representation of $\mathbf{y}$, the output is simply the senone posterior probability conditioned on the input. As in most classification problems, the model can be optimized by minimizing the cross entropy between the senone label and the estimated posterior probability.

When $S$ is greater than $1$, however, it is no longer as simple and direct as in the single-talker case and the label ambiguity or permutation becomes a problem in training. In the case of two speakers, because speech sources are symmetric given the mixture (i.e., $\mathbf{x}_1+\mathbf{x}_2$ equals to $\mathbf{x}_2+\mathbf{x}_1$ and both $\mathbf{x}_1$ and $\mathbf{x}_2$ have the same characteristics), there is no predetermined way to assign the correct target to the corresponding output layer. Interested readers can find additional information in \cite{PIT-yu2017,PIT-Kolbak2017} on how training progresses to nowhere when the conventional supervised approach is used for the multi-talker speech separation.


\section{Permutation Invariant Training for Multi-Talker Speech Recognition} \label{sec:pit}

To address the label ambiguity problem, we propose several architectures based on the permutation invariant training (PIT) \cite{PIT-yu2017,PIT-Kolbak2017} for multi-talker mixed speech recognition. For simplicity and without losing the generality, we always assume there are two-talkers in the mixed speech when describing our architectures in this section.

Note that, DPCL \cite{DeepClustering-hershey2015,DeepClustering2-isik2016} and DANet \cite{AtrractorNet4SpeechSeparation-chen2017} are alternative solutions to the label ambiguity problem when the goal is speech source separation. However, these two techniques cannot be easily applied to direct recognition (i.e., without first separating speech) of multiple streams of speech because of the clustering step required during separation, and the assumption that each time-frequency bin belongs to only one speaker (which is false when the CE criterion is used).

\begin{figure*}[htbp]
  \centering
  \subfigure[Arch\#1: Feature separation with the fixed reference assignment]{
    \includegraphics[width=0.47\linewidth]{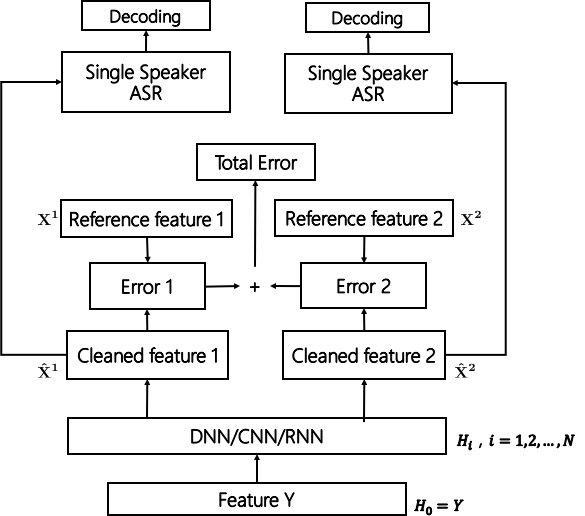}
  }
  ~
  \subfigure[Arch\#2: Feature separation with permutation invariant training]{
    \includegraphics[width=0.47\linewidth]{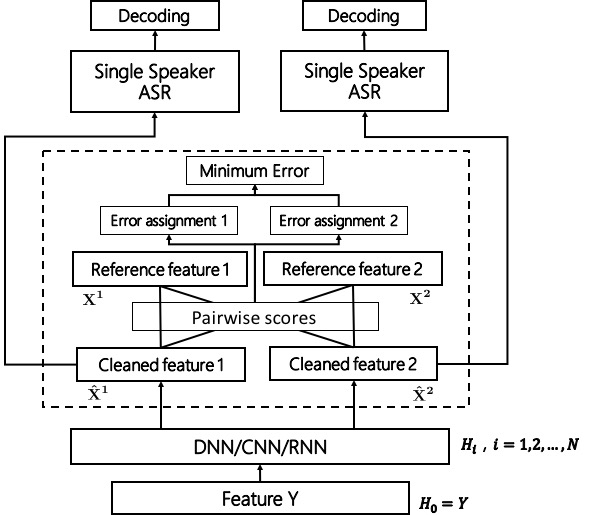}
  }
  \caption{Feature separation architectures for multi-talker mixed speech recognition}
  \label{fig:models1}
\end{figure*}

\subsection{Feature Separation with Direct Supervision}

To recognize the multi-talker mixed speech, one straightforward approach is to estimate the features of each speech source given the mixed speech feature and recognize them one by one using a normal single-talker LVCSR system. This idea is depicted in Figure \ref{fig:models1} where we learn a model to recover the filter bank (FBANK) features from the mixed FBANK features and then feed each stream of the recovered FBANK features to a conventional LVCSR system for recognition.

In the simplest architecture, which is denoted as {\bf{Arch\#1}} and illustrated in Figure \ref{fig:models1}(a), feature separation can be considered as a multi-class regression problem, similar to many previous works \cite{huang2014deep,Weninger:2015:SEL:2965758.2965771,wang2014training,xu2014experimental,huang2015jointtaslp,du2016regression}. In this architecture, $\mathbf{Y}$, the feature of mixed speech, are used as the input to some deep learning models, such as deep neural networks (DNNs), convolutional neural networks (CNNs), and long short-term memory (LSTM) recurrent neural networks (RNNs), to estimate feature representation of each individual talker. If we use the bidirectional LSTM-RNN model, the model will compute
\begin{align}
  \mathbf{H}_0 &= \mathbf{Y} \\
  \mathbf{H}_i^f &= RNN_i^f (\mathbf{H}_{i-1}), i=1, \cdots, N \\
  \mathbf{H}_i^b &= RNN_i^b (\mathbf{H}_{i-1}), i=1, \cdots, N \\
  \mathbf{H_i} &= Stack(H_i^f, H_i^b), i = 1, \cdots, N \\
  \mathbf{\hat{X}}^s &= Linear(\mathbf{H}_{N}), s=1, \cdots, S
\end{align}
where $\mathbf{H}_0$ is the input, $N$ is the number of hidden layers, $\mathbf{H}_i$ is the $i$-th hidden layer, $RNN_i^{f}$ and $RNN_i^{b}$ are the forward and backward RNNs at hidden layer $i$, respectively, $\mathbf{\hat{X}}^s, s=1,\cdots,S$ is the estimated separated features from the output layers for each speech stream $s$. 

During training, we need to provide the correct reference (or target) features $\mathbf{X^s}, s=1,\cdots,S$ for all speakers in the mixed speech to the corresponding output layers for supervision. The model parameters can be optimized to minimize the mean square error (MSE) between the estimated separated feature $\mathbf{\hat X^s}$ and the original reference feature $\mathbf{X^s}$,
\begin{align}
  \mathbf{J} = \frac{1}{S} \min \sum_{s=1}^{S} \sum_t || \mathbf{X_t^s} - \hat{\mathbf{X_t^s}} ||^2
\end{align}
where $S$ is the number of mixed speakers. In this architecture, it is assumed that the reference features are organized in a given order and assigned to the output layer segments accordingly. Once trained, this feature separation module can be used as the front-end to process the mixed speech. The separated feature streams are then fed into a normal single-speaker LVCSR system for decoding.

\subsection{Feature Separation with Permutation Invariant Training}

The architecture depicted in Figure \ref{fig:models1}(a) is easy to implement but with obvious drawbacks. Since the model has multiple output layer segments (one for each stream), and they depend on the same input mixture, assigning reference is actually difficult. The fixed reference order used in this architecture is not quite right since the source speech streams are symmetric and there is no clear clue on how to order them in advance. This is referred to as the label ambiguity (or label permutation) problem in \cite{PIT-yu2017,SingleChannelSep-Weng2015,DeepClustering-hershey2015}. As a result, this architecture may work well on the speaker-dependent setup where the target speaker is known (and thus can be assigned to a specific output segment) during training, but cannot generalize well to the speaker-independent case.

The label ambiguity problem in the multi-talker mixed speech recognition was addressed with limited success in \cite{SingleChannelSep-Weng2015} where Weng et al. assigned reference features depending on the energy level of each speech source. In the architecture illustrated in Figure \ref{fig:models1}(b), named as {\bf{Arch\#2}}, permutation invariant training (PIT) \cite{PIT-yu2017,PIT-Kolbak2017} is utilized to estimate individual feature streams. In this architecture, The reference feature sources are given as a set instead of an ordered list. The output-reference assignment is determined dynamically based on the current model. More specifically, it first computes the MSE for each possible assignment between the reference $\mathbf{X^{s'}}$ and the estimated source $\mathbf{\hat X^{s}}$, and picks the one with minimum MSE. In other words, the training criterion is 
\begin{align}
  J = \frac{1}{S} \min_{s' \in permu(S)} \sum_s \sum_t || \mathbf{X_t^{s'}} - \mathbf{\hat{X}_t^s} ||^2, s=1, \cdots, S
\end{align}
where $permu(S)$ is a permutation of $1, \cdots, S$. We note two important ingredients in this objective function. First, it automatically finds the appropriate assignment no matter how the labels are ordered. Second, the MSE is computed over the whole sequence for each assignment. This forces all the frames of the same speaker to be aligned with the same output segment, which can be regarded as performing the feature-level tracing implicitly. With this new objective function, We can simultaneously perform label assignment and error evaluation on the feature level. It is expected that the feature streams separated with PIT (Figure \ref{fig:models1}(b)) has higher quality than that separated with fixed reference order (Figure \ref{fig:models1}(a)). As a result, the recognition errors on these feature streams should also be lower. Note that the computational cost associated with permutation is negligible compared to the network forward computation during training, and no permutation (and thus no cost) is needed during evaluation. 

\begin{figure*}[!htbp]
  \centering
    \subfigure[Arch\#3: Direct multi-talker mixed speech recognition with PIT]{
    \includegraphics[width=0.47\linewidth]{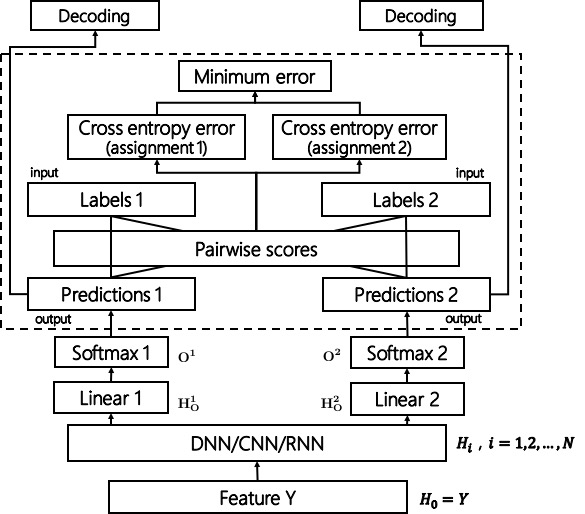}
  }
\subfigure[Arch\#4: Joint optimization of PIT-based feature separation and recognition]{
    \includegraphics[width=0.47\linewidth]{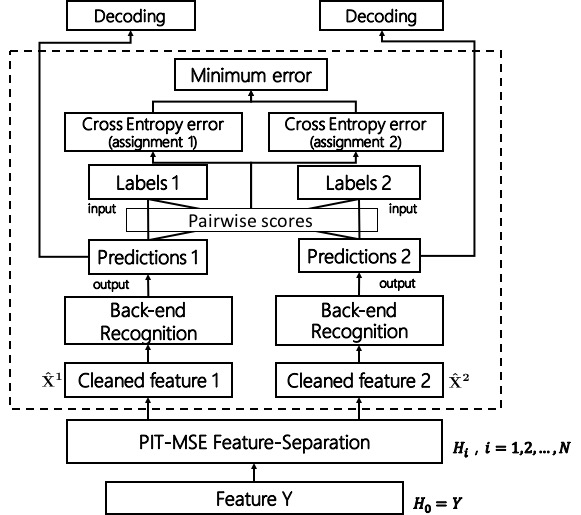}
  }
  \caption{Advanced architectures for multi-talker mixed speech recognition}
  \label{fig:models2}
\end{figure*}

\subsection{Direct Multi-Talker Mixed Speech Recognition with PIT}

In the previous two architectures mixed speech features are first separated explicitly and then recognized independently with a conventional single-talker LVCSR system. Since the feature separation is not perfect, there is mismatch between the separated features and the normal features used to train the conventional LVCSR system. In addition, the objective function of minimizing the MSE between the estimated and reference features is not directly related to the recognition performance. In this section, we propose an end-to-end architecture that directly recognizes mixed speech of multiple speakers.

In this architecture, denoted as {\bf{Arch\#3}}, we apply PIT to the CE between the reference and estimated senone posterior probability distributions as shown in Figure \ref{fig:models2}(a). Given some feature representation $\mathbf{Y}$ of the mixed speech $\mathbf{y}$, this model will compute 
\begin{align}
  \mathbf{H}_0 &= \mathbf{Y} \\
  \mathbf{H}_i^{f} &= RNN_i^{f}(\mathbf{H}_{i-1}), i=1,\cdots,N \\
  \mathbf{H}_i^{b} &= RNN_i^{b}(\mathbf{H}_{i-1}), i=1,\cdots,N \\
  \mathbf{H}_i &= Stack(\mathbf{H}_i^{f}, \mathbf{H}_i^{b}), i=1,\cdots,N \\
  \mathbf{H}_o^s &= Linear(\mathbf{H}_N), s=1,\cdots,S \\
  \mathbf{O}^s &= Softmax(\mathbf{H}_o^s), s=1,\cdots,S
\end{align}
using a deep bidirectional RNN, where Equations (8)$\sim$(11) are similar to Equations (1)$\sim$(4). $\mathbf{H}_o^s, s=1,\cdots,S$ is the excitation at output layer for each speech stream $s$, and $\mathbf{O}^s, s=1,\cdots,S$ is the output segment for stream $s$. Different from architectures discussed in previous sections, in this architecture each output segment represents the estimated senone posterior probability for a speech stream. No additional feature separation, clustering or speaker tracing is needed. Although various neural network structures can be used, in this study we focus on bidirectional LSTM-RNNs. 

In this direct multi-talker mixed speech recognition architecture, we minimize the objective function

\begin{align}
  J &= \frac{1}{S} \min_{s' \in permu(S)} \sum_{s}  \sum_{t} { CE(\mathbf{\ell}_t^{s'},\mathbf{O}_t^s)}, s=1,\cdots,S
\end{align}

In other words, we minimize the minimum average CE of every possible output-label assignment. All the frames of the same speaker are forced to be aligned with the same output segment by computing the CE over the whole sequence for each assignment. This strategy allows for the direct multi-talker mixed speech recognition without explicit separation. It is a simpler and more compact architecture for multi-talker speech recognition. 

\subsection{Joint Optimization of PIT-based Feature Separation and Recognition}

As mentioned above, the main drawback of the feature separation architectures is the mismatch between the distorted separation result and the features used to train the single-talker LVCSR system. The direct multi-talker mixed speech recognition with PIT, which bypassed the feature separation step, is one solution to this problem. Here we propose another architecture named joint optimization of PIT-based feature separation and recognition, and it is denoted as {\bf{Arch\#4}} and shown in Figure \ref{fig:models2}(b).

This architecture contains two PIT-components, the front-end feature separation module with PIT-MSE and the back-end recognition module with PIT-CE. Different from the architecture in Figure \ref{fig:models1}(b), in this architecture a new LVCSR system is trained upon the output of the feature separation module with PIT-CE. The whole model is trained progressively: the front-end feature separation module is firstly optimized with PIT-MSE; Then the parameters in the back-end recognition module are optimized with PIT-CE while keeping the parameters in the feature separation module fixed. Finally parameters in both modules are jointly refined with PIT-CE using a small learning rate. Note that the reference assignment in the recognition (PIT-CE) step is the same as that in the separation (PIT-MSE) step.

\begin{align}
  J_{1} = \frac{1}{S} \min_{s' \in permu(S)} \sum_s \sum_t || \mathbf{X_t^{s'}} - \mathbf{\hat{X}_t^s} ||^2, s=1, \cdots, S
\end{align}

\begin{align}
  J_2 &= \frac{1}{S} \min_{s' \in permu(S)} \sum_{s}  \sum_{t} { CE(\mathbf{\ell}_t^{s'},\mathbf{O}_t^s)}, s=1,\cdots,S
\end{align}

During decoding, the mixed speech features are fed into this architecture, and the final posterior streams are used for decoding as normal.

\section{Experimental Results} \label{sec:exp}

To evaluate the performance of the proposed architectures, we conducted a series of experiments on an artificially generated two- and three-talker mixed speech dataset based on the AMI corpus \cite{hain2012transcribing}.

\def \fig {figure/}

There are four reasons for us to use AMI: 1) AMI is a speaker-independent spontaneous LVCSR corpora. Compared to small vocabulary, speaker-dependent, read English datasets used in most of the previous studies \cite{MonauralSpeechSepChallenge-Cooke2010,SingleChannelSep-Weng2015,HERSHEY201045,rennie2010single}, observations made and conclusions drawn from AMI are more likely generalized to other real-world scenarios; 2) AMI is a really hard task with different kinds of noises, truly spontaneous meeting style speech, and strong accents. It reflects the true ability of LVCSR when the training set size is around 100hr. The state-of-the-art word error rate (WER) on AMI is around 25.0\% for the close-talk condition \cite{PureMMI-Povey2016} and more than 45.0\% for the far-field condition with single-microphone \cite{PureMMI-Povey2016,HighwayBLSTM-zhang2016}. These WERs are much higher than that on other corpora, such as Switchboard \cite{godfrey-switchboard} on which the WER is now below 10.0\% \cite{HumanParity-Xiong2016,PureMMI-Povey2016,CNN-Dilated-sercu2016,IBM-SWB-Saon2016}; 3) Although the close-talk data (AMI IHM) was used to generate mixed speech in this work, the existence of parallel far-field data (AMI SDM/MDM) allows us to evaluate our architectures based on the far-field data in the future; 4) AMI is a public corpora, using AMI allows interested readers to reproduce our results more easily.

The AMI IHM (close-talk) dataset contains about 80hr and 8hr speech in training and evaluation sets, respectively \cite{hain2012transcribing,swietojanski2013hybrid}. Using AMI IHM, we generated a two-talker (IHM-2mix) and a three-talker (IHM-3mix) mixed speech dataset. 

To artificially synthesize IHM-2mix, we randomly select two speakers and then randomly select an utterance for each speaker to form a mixed-speech utterance. For easier explanation, the high energy (High E) speaker in the mixed speech is always chosen as the target speaker and the low energy (Low E) speaker is considered as interference speaker. We synthesized mixed speech for five different SNR conditions (i.e. 0dB, 5dB, 10dB, 15dB, 20dB) based on the energy ratio of the two-talkers. To eliminate easy cases we force the lengths of the selected source utterances comparable so that at least half of the mixed speech contains overlapping speech. When the two source utterances have different lengths, the shorter one is padded with small noise at the front and end. The same procedure is used for preparing both the training and testing data. We generated in total 400hr two-talker mixed speech, 80hr per SNR condition, as the training set. A subset of 80hr speech from this 400hr training set was used for fast model training and evaluation. For evaluation, total 40hr two-talker mixed speech, 8hr per SNR condition, is generated and used.

The IHM-3mix dataset was generated similarly. The relative energy of the three speakers in each mixed utterance varies randomly in the training set. Different from the training set, all the speakers in the same mixed utterance have equal energy in the testing set. We generated in total 400hr and 8hr three-talker mixed speech as the training and testing set, respectively. 

 \begin{figure}
   \centering   \includegraphics[width=0.9\linewidth]{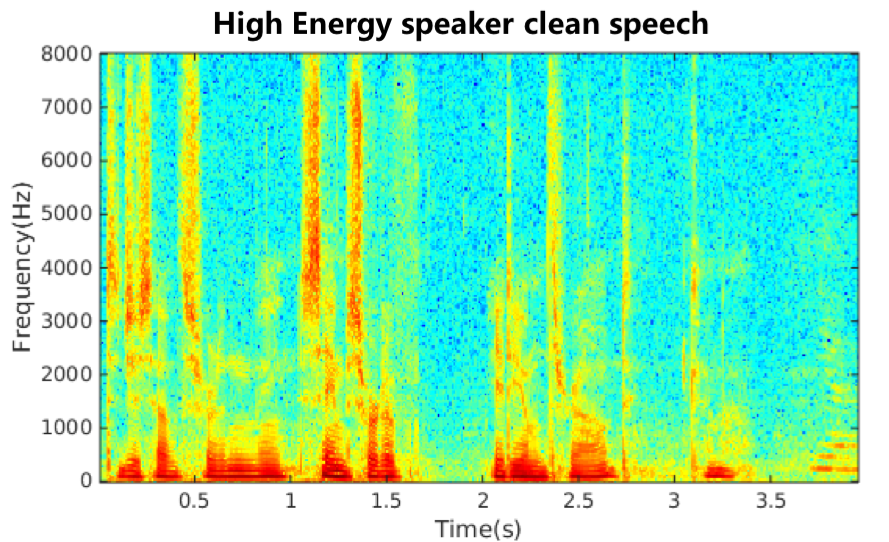}   \includegraphics[width=0.9\linewidth]{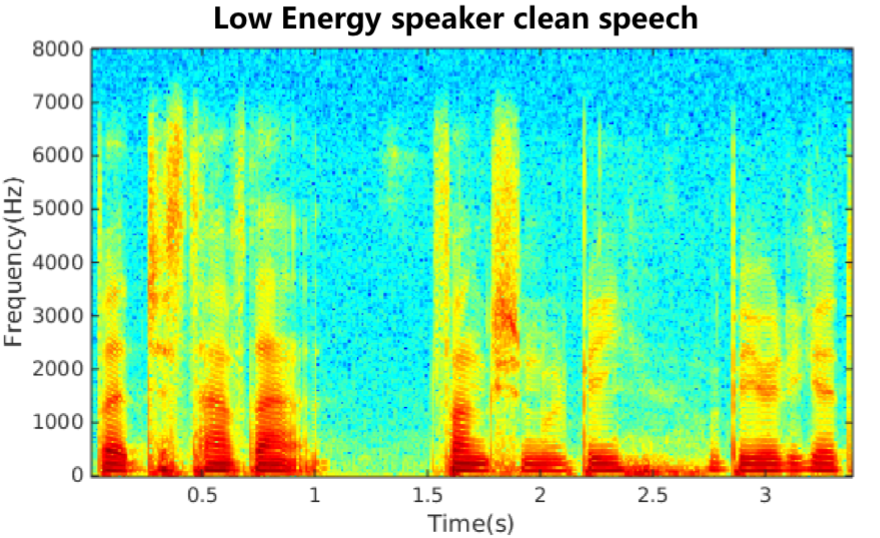}   \includegraphics[width=0.9\linewidth]{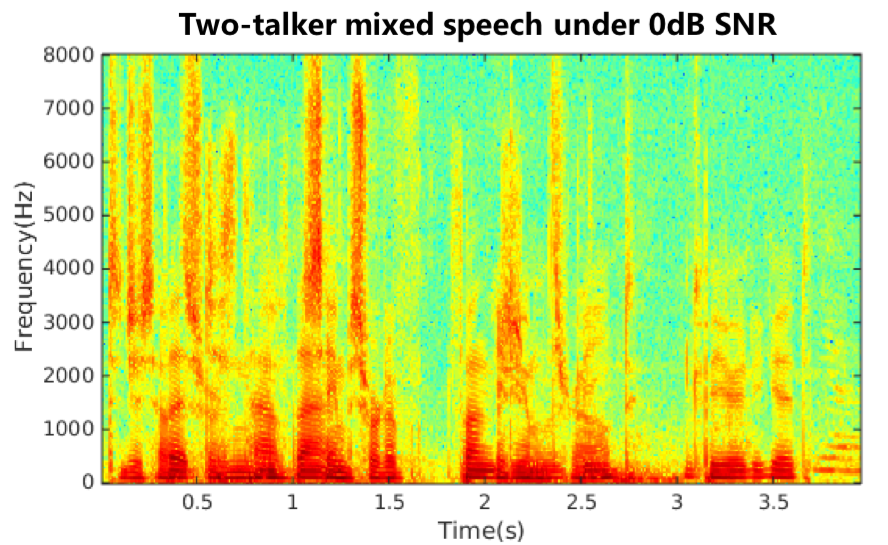}    
     \caption{Spectrogram comparison between the original single-talker clean speech and the 0db two-talker mixed-speech in the IHM-2mix dataset}
     \label{fig:spectrum}
 \end{figure}

Figure \ref{fig:spectrum} compares the spectrogram of a single-talker clean utterance and the corresponding 0db two-talker mixed utterance in the IHM-2mix dataset. Obviously it is really hard to separate the spectrogram and reconstruct the source utterances by visually examining it. 

\subsection{Single-speaker Recognition Baseline}

In this work, all the neural networks were built using the latest Microsoft Cognitive Toolkit (CNTK) \cite{yu2014introduction} and the decoding systems were built based on Kaldi \cite{povey2011kaldi}. We first followed the officially released kaldi recipe to build an LDA-MLLT-SAT GMM-HMM model. This model uses 39-dim MFCC feature and has roughly 4K tied-states and 80K Gaussians. We then used this acoustic model to generate the senone alignment for neural network training. We trained the DNN and BLSTM-RNN baseline systems with the original AMI IHM data. 80-dimensional log filter bank (LFBK) features with CMVN were used to train the baselines. The DNN has 6 hidden layers each of which contains 2048 Sigmoid neurons. The input feature for DNN contains a window of 11 frames. The BLSTM-RNN has 3 bidirectional LSTM layers which are followed by the softmax layer. Each BLSTM layer has 512 memory cells. The input to the BLSTM-RNN is a single acoustic frame. All the models explored here are optimized with cross-entropy criterion. The DNN is optimized using SGD method with 256 minibatch size, and the BLSTM-RNN is trained using SGD with 4 full-length utterances in each minibatch.

For decoding, we used a 50K-word dictionary and a trigram language model interpolated from the ones created using the AMI transcripts and the Fisher English corpus. The performance of these two baselines on the original single-speaker AMI corpus are presented in Table \ref{tab:baselinewerami}. These results are comparable with that reported by others \cite{swietojanski2013hybrid} even though we did not use adapted fMLLR feature. It is noted that adding more BLSTM layers did not show meaningful WER reduction in the baseline.

\begin{table}[th]
  \caption{WER (\%) of the baseline systems on original AMI IHM single-talker corpus}
  \label{tab:baselinewerami}
  \centering
  \begin{tabular}{ c  c  c }
    \toprule
    \textbf{Model} & \textbf{WER}  \\
    \midrule
    DNN            & 28.0             \\
    BLSTM          & 26.6               \\
    \bottomrule
  \end{tabular}
  
\end{table}

To test the normal single-speaker model on the two-talker mixed speech, the above baseline BLSTM-RNN model is utilized to decode the mixed speech directly. During scoring we compare the decoding output (only one output) with the reference of each source utterance to obtain the WER for the corresponding source utterance. Table \ref{tab:baselinewermixed} summarizes the recognition results. It is clear, from the table, that the single-speaker model performs very poorly on the multi-talker mixed speech as indicated by the huge WER degradation of the high-energy speaker when SNR decreases. Further more, in all the conditions, the WERs for the low energy speaker are all above 100.0\%. These results demonstrate the great challenge in the multi-talker mixed speech recognition.

\begin{table}[th]
  \caption{WER (\%) of the baseline BLSTM-RNN single-speaker system on the IHM-2mix dataset}
  \label{tab:baselinewermixed}
  \centering
  \begin{tabular}{ c  c  c }
    \toprule
    \textbf{SNR Condition} & \textbf{High E Spk} & \textbf{Low E Spk}  \\
    \midrule
    0db           & 85.0  & 100.5               \\
    5db           & 68.8  & 110.2             \\
    10db           & 51.9  & 114.9               \\
    15db           & 39.3  & 117.6               \\
    20db           & 32.1  & 118.7               \\
    \bottomrule
  \end{tabular}
  
\end{table}

\subsection{Evaluation of Two-talker Speech Recognition Architectures}

The proposed four architectures for two-taker speech recognition are evaluated here. For the first two approaches (Arch\#1 and Arch\#2) that contain an explicit feature separation stage (with and without PIT-MSE), a 3-layer BLSTM is used in the feature separation module. The separated feature streams are fed into a normal 3-layer BLSTM LVCSR system, trained with single-talker speech, for decoding. The whole system contains in total six BLSTM layers. For the other two approaches (Arch\#3 and Arch\#4), in which PIT-CE is used, 6-layer BLSTM models are used so that the number of parameters is comparable to the other two architectures. In all these architectures the input is the 40-dimensional LFBK feature and each layer contains 768 memory cells. To train the latter two architectures that exploit PIT-CE we need to prepare the alignments for the mixed speech. The senone alignments for the two-talkers in each mixed speech utterance are from the single-speaker baseline alignment. The alignment of the shorter utterance within the mixed speech is padded with the silence state at the front and the end. All the models were trained with a minibatch of 8 utterances. The gradient was clipped to 0.0003 to guarantee the training stability. To obtain the results reported in this section we used the 80hr mixed speech training subset. 

The recognition results on both speakers are evaluated. For scoring, we evaluated the two hypotheses, obtained from two output sections, against the two references and pick the assignment with better WER to compute the final WER.

The results on the 0db SNR condition are shown in Table \ref{tab:diff_pitwer_nn}. Compared to the 0dB condition in Table \ref{tab:baselinewermixed}, all the proposed multi-talker speech recognition architectures obtain obvious improvement on both speakers. Within the two architectures with the explicit feature separation stage, the architecture with PIT-MSE is significantly better than the baseline feature separation architecture. These results confirmed that the label permutation problem can be well alleviated by the PIT-MSE at the feature level. We can also observe that applying PIT-CE on the recognition module (Arch\#3 \& Arch\#4) can further reduce WER by 10.0\% absolute. This is because these two architectures can significantly reduce the mismatch between the separated feature and the feature used to train the LVCSR model. It is also because cross-entropy is more directly related to the recognition accuracy. Comparing Arch\#3 and Arch\#4, we can see that the architecture with joint optimization on PIT-based feature separation and recognition slightly outperforms the direct PIT-CE based model.

Since Arch\#3 and Arch\#4 achieve comparable results, and the model architecture and training process of Arch\#3 is much simpler than that of Arch\#4, our further evaluations reported in the following sections are based on Arch\#3. For clarity, Arch\#3 is named {\bf{direct PIT-CE-ASR}} from now on.

\begin{table}[th]
  \caption{WER (\%) of the proposed multi-talker mixed speech recognition architectures on the IHM-2mix dataset under 0db SNR condition (using 80hr training subset). Arch\#1-\#4 indicate the proposed architectures described in Section III.A-D, respectively}
  \label{tab:diff_pitwer_nn}
  \centering
  \begin{tabular}{c c c |  c  c }
    \toprule
    \textbf{Arch} & \textbf{Front-end} & \textbf{Back-end} & \textbf{High E WER} & \textbf{Low E WER}  \\
    \midrule
    \#1 & Feat-Sep-baseline & Single-Spk-ASR  &  72.58 & 79.61 \\
    \#2 & Feat-Sep-PIT-MSE & Single-Spk-ASR & 68.88 & 75.62 \\
     \midrule
     \#3  & $\times$ & PIT-CE & 59.72 & 66.96 \\
     \#4  & Feat-Sep-PIT-MSE & PIT-CE & 58.68 & 66.25 \\
    \bottomrule
  \end{tabular}
\end{table}

\subsection{Evaluation of the Direct PIT-CE-ASR Model on Large Dataset}

\begin{figure*}[!htbp]
  \centering
  \includegraphics[width=\linewidth]{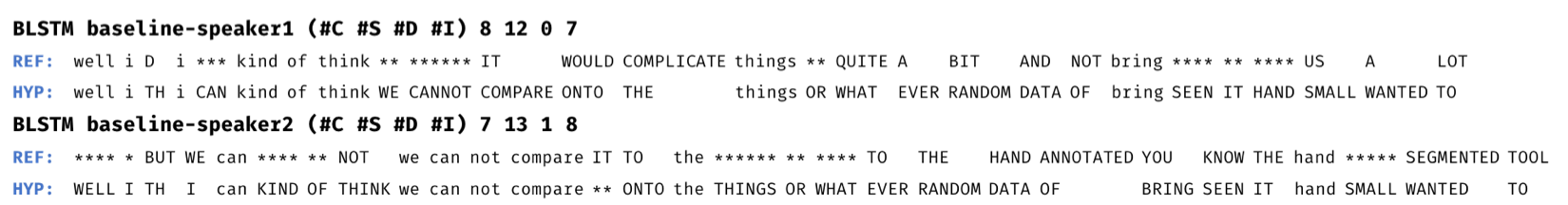}
  \caption{Decoding results of baseline single speaker BLSTM-RNN system on 0db two-talker mixed speech sample}
  \label{fig:decodeSampleBs}
\end{figure*}
\begin{figure*}[!htbp]
  \includegraphics[width=\linewidth]{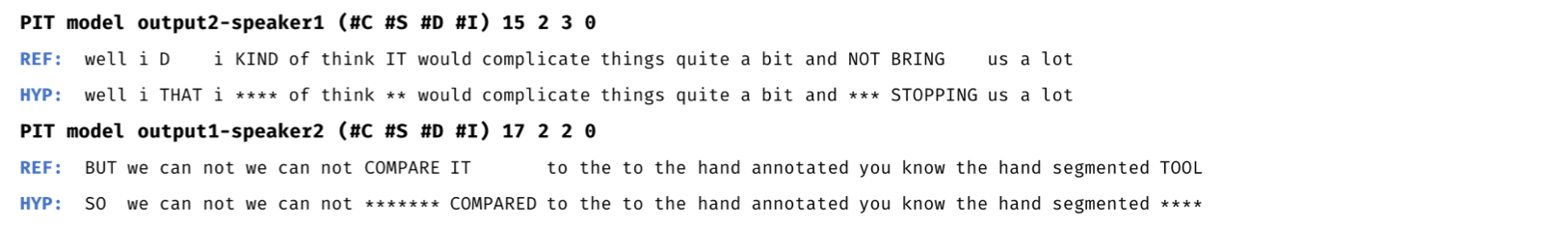}
  \caption{Decoding results of the proposed direct PIT-CE-ASR model on 0db two-talker mixed speech sample}
  \label{fig:decodeSamplePIT}
\end{figure*}

We evaluated the direct PIT-CE-ASR architecture on the full IHM-2mix corpus. All the 400hr mixed data under different SNR conditions are pooled together for training. The direct PIT-CE-ASR model is still composed of 6 BLSTM layers with 768 memory cells in each layer. All other configurations are also the same as the experiments conducted on the subset.

The results under different SNR conditions are shown in Table \ref{tab:pitwer}. The direct PIT-CE-ASR model achieved significant improvements on both talkers compared to baseline results in Table \ref{tab:baselinewermixed} for all SNR conditions. Comparing to the results in Table \ref{tab:diff_pitwer_nn}, achieved with 80hr training subset, we observe that additional absolute 10.0\% WER improvement on both speakers can be obtained using the large training set. We also observe that the WER increases slowly when the SNR becomes smaller for the high energy speaker, and the WER improvement is very significant for the low energy speaker across all conditions. In the 0dB SNR scenario, the WERs on two speakers are very close and are 45.0\% less than that achieved with the single-talker ASR system for both high and low energy speakers. At 20dB SNR, the WER of the high energy speaker is still significantly better than the baseline, and approaches the single-talker recognition result reported in Table \ref{tab:baselinewerami}.  

\begin{table}[th]
  \caption{WER (\%) of the proposed direct PIT-CE-ASR model on the IHM-2mix dataset with full training set}
  \label{tab:pitwer}
  \centering
  \begin{tabular}{c  c  c }
    \toprule
    \textbf{SNR Condition} & \textbf{High E WER} & \textbf{Low E WER}  \\
    \midrule
    0db & 47.77  & 54.89 \\
    5db & 39.25  & 59.24  \\
    10db & 33.83  & 64.14 \\
    15db & 30.54  & 71.75 \\
    20db & 28.75  & 79.88 \\
    \bottomrule
  \end{tabular}
\end{table}

\subsection{Permutation Invariant Training with Alternative Deep Learning Models}

We investigated the direct PIT-CE-ASR model with alternative deep learning models. The first model we evaluated is a 6-layer feed-forward DNN in which each layer contains 2048 Sigmoid units. The input to the DNN is a window of 11 frames each with a 40-dimensional LFBK feature.

The results of DNN-based PIT-CE-ASR model is reported at the top of Table \ref{tab:pitwer-diff-nn}. Although it still gets obvious improvement over the baseline single-speaker model, the gain is much smaller with near 20.0\% WER difference in every condition than that from BLSTM-based PIT-CE-ASR model. The difference between DNN and BLSTM models partially attribute to the stronger modeling power of BLSTM models and partially attribute to the better tracing ability of RNNs.

We also compared the BLSTM models with 4, 6, and 8 layers as shown in Table \ref{tab:pitwer-diff-nn}. It is observed that deeper BLSTM models perform better. This is different from the single speaker ASR model whose performance peaks at 4 BLSTM layers \cite{HighwayBLSTM-zhang2016}. This is because the direct PIT-CE-ASR architecture needs to conduct two tasks - separation and recognition, and thus requires additional modeling power.

\begin{table}[th]
  \caption{WER (\%) of the direct PIT-CE-ASR model using different deep learning models on the IHM-2mix dataset}
  \label{tab:pitwer-diff-nn}
  \centering
  \begin{tabular}{c | c  c  c }
    \toprule
    \textbf{Models} & \textbf{SNR Condition} & \textbf{High E WER} & \textbf{Low E WER}  \\
    \midrule
    \multirow{5}{*}{6L-DNN} & 0db & 72.95 & 80.29 \\
     & 5db & 65.42 & 84.44 \\
     & 10db & 55.27 & 86.55 \\
     & 15db & 47.12 & 89.21 \\
     & 20db & 40.31 & 92.45 \\
    \midrule
    \multirow{5}{*}{4L-BLSTM} & 0db & 49.74  & 56.88 \\
     & 5db & 40.31  & 60.31 \\
     & 10db & 34.38  & 65.52 \\
     & 15db & 31.24  & 73.04 \\
     & 20db & 29.68  & 80.83 \\
    \midrule
    \multirow{5}{*}{6L-BLSTM} & 0db & 47.77  & 54.89 \\
     & 5db & 39.25  & 59.24  \\
     & 10db & 33.83  & 64.14 \\
     & 15db & 30.54  & 71.75 \\
     & 20db & 28.75  & 79.88 \\
    \midrule
    \multirow{5}{*}{8L-BLSTM} & 0db & 46.91	& 53.89 \\
     & 5db & 39.14 & 59.00 \\
     & 10db & 33.47 & 63.91 \\
     & 15db & 30.09	& 71.14 \\
     & 20db & 28.61	& 79.34 \\
    \bottomrule
  \end{tabular}
\end{table}


\subsection{Analysis on Multi-Talker Speech Recognition Results}

To better understand the results on multi-talker speech recognition, we computed the WER separately for the speech mixed with same and opposite genders. The results are shown in Table \ref{tab:pitwer_gender}. It is observed that the same-gender mixed speech is much more difficult to recognize than the opposite-gender mixed speech, and the gap is even larger when the energy ratio of the two speakers is closer to 1. It is also observed that the mixed speech of two male speakers is hard to recognize than that of two female speakers. These results suggest that effective exploitation of gender information may help to further improve the multi-talker speech recognition system. We will explore this in our future work.

\begin{table}[th]
  \caption{WER (\%) comparison of the 6-layer-BLSTM direct PIT-CE-ASR model on the mixed speech generated from two male speakers ({\bf M + M}), two female speakers ({\bf F + F}) and a male and a female speaker ({\bf M + F})}
  \label{tab:pitwer_gender}
  \centering
  \begin{tabular}{c | c  c  c }
    \toprule
    \textbf{Genders} & \textbf{SNR Condition} & \textbf{High E WER} & \textbf{Low E WER}  \\
    \midrule
    \multirow{3}{*}{M + M} & 0db & 52.18 & 59.32 \\
     & 5db & 42.64 & 61.77 \\
     & 10db & 36.10 & 63.94\\
    \midrule
    \multirow{3}{*}{F + F} & 0db & 49.90 & 57.59 \\
     & 5db & 40.02 & 60.92 \\
     & 10db & 32.47 & 65.15\\
    \midrule
    \multirow{3}{*}{M + F} & 0db & 44.89 & 51.72 \\
     & 5db & 37.34 & 57.43 \\
     & 10db & 33.22 & 63.86 \\
    \bottomrule
  \end{tabular}
\end{table}

To further understand our model, we examined the recognition results with and without using the direct PIT-CE-ASR. An example of these results on a 0db two-talker mixed speech utterance is shown in Figure \ref{fig:decodeSampleBs} (using the single-speaker baseline system) and \ref{fig:decodeSamplePIT} (with direct PIT-CE-ASR). It is clearly seen that the results are  erroneous when the single-speaker baseline system is used to recognize the two-talker mixed speech. In contrast, much more words are recognized correctly with the proposed direct PIT-CE-ASR model. 

\subsection{Three-Talker Speech Recognition with Direct PIT-CE-ASR}

In this subsection, we further extend and evaluate the proposed direct PIT-CE-ASR model on the three-talker mixed speech using the IHM-3mix dataset.

\begin{figure}
  \centering   \includegraphics[width=0.9\linewidth]{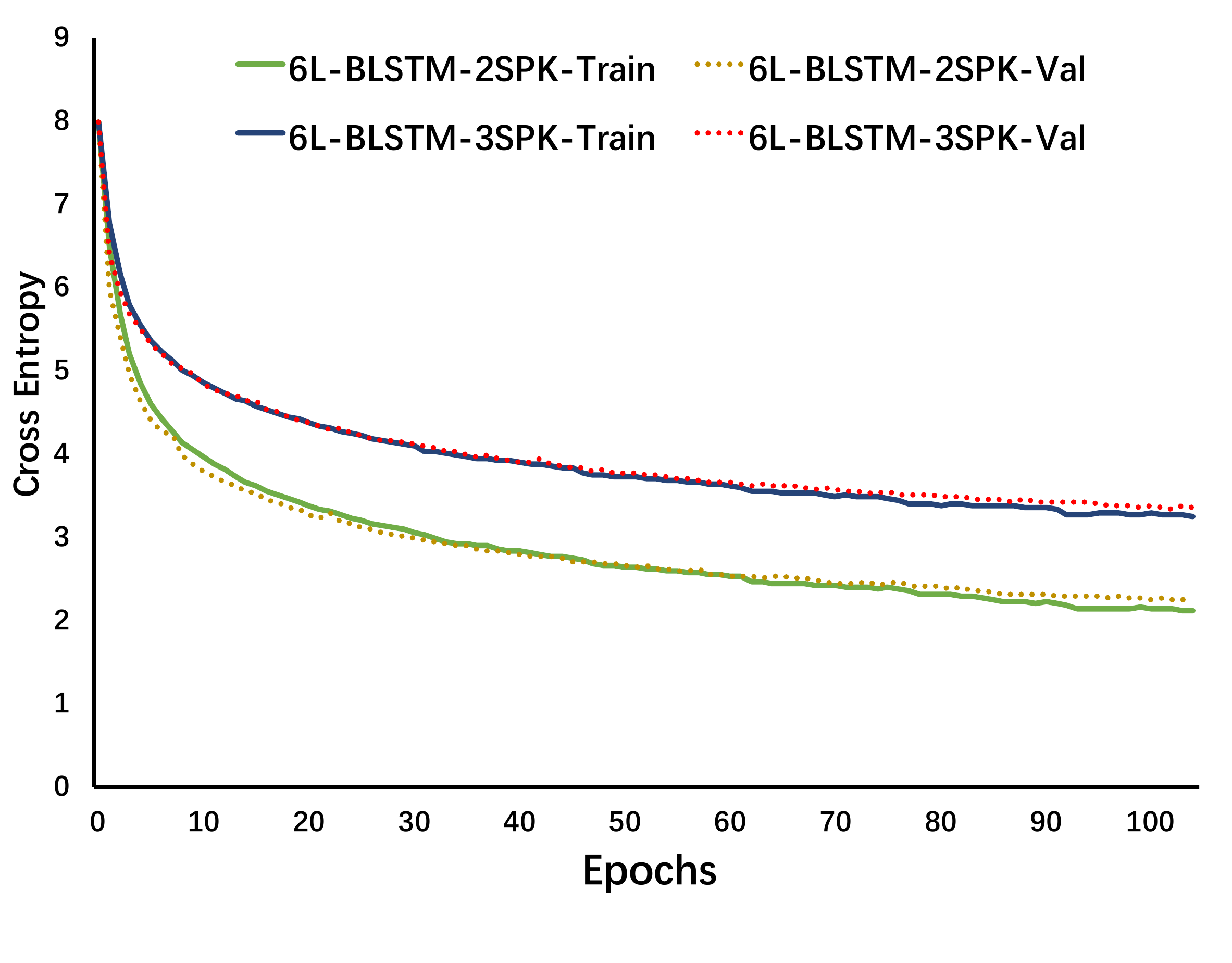}
  \caption{CE values over epochs on both the IHM-2mix and IHM-3mix training and validation sets with the proposed direct PIT-CE-ASR model}
  \label{fig:figure_ce}
\end{figure}

The three-talker direct PIT-CE-ASR model is also a 6-layer BLSTM model. The training and testing configurations are the same as those for two-talker speech recognition. The direct PIT-CE-ASR training processes as measured by CE on both two- and three-talker mixed speech training and validation sets are illustrated in Figure \ref{fig:figure_ce}. It is observed that the direct PIT-CE-ASR model with this specific configuration converges slowly, and the CE improvement progress on the training and validation sets is almost the same. The training progress on three-talker mixed speech is similar to that on two-talker mixed speech, but with an obviously higher CE value. This indicates the huge challenge when recognizing speech mixed with more than two talkers. Note that, in this set of experiments we used the same model configuration as that used in two-talker mixed speech recognition. Since three-talker mixed speech recognition is much harder, using deeper and wider models may help to improve performance. Due to resource limitation, we did not search for the best configuration for the task.

The three-talker mixed speech recognition WERs are reported in Table \ref{tab:three-spk}. The WERs on different gender combinations are also provided. The WERs achieved with the single-speaker model are listed at the first line in Table \ref{tab:three-spk}. Compared to the results on IHM-2mix, the results on IHM-3mix are significantly worse using the conventional single speaker model. Under this extremely hard setup, the proposed direct PIT-CE-ASR architecture still demonstrated its powerful ability on separating/tracing/recognizing the mixed speech, and achieved 25.0\% relative WER reduction across all three speakers. Although the performance gap from two-talker to three-talker is obvious, it is still very promising under this speaker-independent three-talker LVCSR task. Not surprisingly, the mixed speech of different genders is relatively easier to recognize than that of same gender.

\begin{table}[th]
  \caption{WER (\%) comparison of the baseline single-speaker BLSTM-RNN system and the proposed direct PIT-CE-ASR model on the IHM-3mix dataset. {\bf Diff} indicates the mixed speech is from different genders, and {\bf Same} indicates the mixed speech is from same gender}
  \label{tab:three-spk}
  \centering
  \begin{tabular}{c  c c  c c}
    \toprule
    \textbf{Genders} & \textbf{Model} & \textbf{Speaker1} & \textbf{Speaker2} & \textbf{Speaker3} \\
    \midrule
    All     &  BLSTM-RNN    & 91.0 & 90.5 & 90.8 \\
    \midrule
    All     &   \multirow{3}{*}{direct PIT-CE-ASR}   & 69.54 & 67.35 & 66.01   \\
    Different     &    & 69.36 & 65.84 & 64.80    \\
    Same      &    & 72.21 & 70.11 & 69.78     \\
    \bottomrule
  \end{tabular}
\end{table}


Moreover, we conducted another interesting experiment. We used the three-talker PIT-CE-ASR model to recognize the two-talker mixed speech. The results are shown in Table \ref{tab:pitwer_three2two}. Surprisingly, the results are almost identical to that obtained using the 6-layer BLSTM based two-talker model (shown in Table \ref{tab:pitwer}). This demonstrates the good generalization ability of our proposed direct PIT-CE-ASR model over variable number of mixed speakers. This suggests that a single PIT model may be able to recognize mixed speech of different number of speakers without knowing or estimating the number of speakers.

\begin{table}[th]
  \caption{WER (\%) of using three-talker direct PIT-CE-ASR model to recognize two-talker mixed IHM-2mix speech}
  \label{tab:pitwer_three2two}
  \centering
  \begin{tabular}{c | c  c  c }
    \toprule
    \textbf{Model} & \textbf{SNR Condition} & \textbf{High E WER} & \textbf{Low E WER} \\
    \midrule
    \multirow{5}{*}{Three-Talker PIT-CE-ASR} & 0db           & 46.63 & 54.59  \\
     & 5db           & 39.47 & 59.78  \\
     & 10db           & 34.50 & 64.55  \\
     & 15db           & 32.03 & 72.88  \\
     & 20db           & 30.66 & 81.63  \\
    \bottomrule
  \end{tabular}
\end{table}

\section{Conclusion} \label{sec:conclusion}

In this paper, we proposed several architectures for recognizing multi-talker mixed speech given only a single channel of the mixed signal. Our technique is based on permutation invariant training, which was originally developed for separation of multiple speech streams. PIT can be performed on the front-end feature separation module to obtain better separated feature streams or be extended on the back-end recognition module to predict the separated senone posterior probabilities directly. Moreover, PIT can be implemented on both front-end and back-end with a joint-optimization architecture. When using PIT to optimize a model, the criterion is computed over all frames in the whole utterance for each possible output-target assignment, and the one with the minimum loss is picked for parameter optimization. Thus PIT can address the label permutation problem well, and conduct the speaker separation and tracing in one shot. Particularly for the proposed architecture with the direct PIT-CE based recognition model, multi-talker mixed speech recognition can be directly conducted without an explicit separation stage. 

The proposed architectures were evaluated and compared on an artificially mixed AMI dataset with both two- and three-talker mixed speech. The experimental results indicate that the proposed architectures are very promising. Our models can obtain relative 45.0\% and 25.0\% WER reduction against the state-of-the-art single-talker speech recognition system across all speakers when their energies are comparable, for two- and three-talker mixed speech, respectively. Another interesting observation is that there is even no degradation when using proposed three-talker model to recognize the two-talker mixed speech directly. This suggests that we can construct one model to recognize speech mixed with variable number of speakers without knowing or estimating the number of speakers in the mixed speech. To our knowledge, this is the first work on the multi-talker mixed speech recognition on the  challenging speaker-independent spontaneous LVCSR task.


\section*{Acknowledgment}

This work was supported by the Shanghai Sailing Program No. 16YF1405300, the China NSFC projects (No. 61573241 and No. 61603252), the Interdisciplinary Program (14JCZ03) of Shanghai Jiao Tong University in China, and the Tencent-Shanghai Jiao Tong University joint project. Experiments have been carried out on the PI supercomputer at Shanghai Jiao Tong University.

\ifCLASSOPTIONcaptionsoff
  \newpage
\fi

\bibliographystyle{IEEEtran}
\bibliography{IEEEabrv,mybib}

%





\end{document}